# NASA Operational Simulator for Small Satellites (NOS$^3$): the STF-1 CubeSat case study
[Integration & Testing; Flight Software]

Dustin M. Geletko, Matthew D. Grubb, John P. Lucas, Justin R. Morris, Max Spolaor, Mark D. Suder, Steven C. Yokum, and Scott A. Zemerick.
*NASA Independent Verification and Validation (IV&V), Independent Test Capability (ITC)*
*Jon McBride Software Testing & Research (JSTAR) Laboratory*
*Fairmont, West Virginia 26554, U.S.A.*

Max Spolaor – max.spolaor@nasa.gov (for publication)

**ABSTRACT**

One of the primary objectives of small satellites is to reduce the costs associated with spacecraft development and operations as compared to traditional spacecraft missions. Small satellite missions are generally able to reduce mission planning, hardware, integration, and operational costs; however, small satellite missions struggle with reducing software development and testing costs. This paper presents the case study of the NASA Operational Simulator for Small Satellites (NOS$^3$), a software-only simulation framework that was developed for the Simulation-to-Flight 1 (STF-1) 3U CubeSat mission. The general approach is to develop software simulators for the various hardware flight components (e.g., electrical power system, antenna deployment system, etc.) to create a completely virtual representation of the actual spacecraft system. In addition, NOS$^3$ conveniently packages together a set of open-source software packages including the "42" dynamics simulator, the spacecraft software development framework (core Flight System), and a command and control system (COSMOS). This results in a flexible and easily deployable simulation environment that can be utilized to support software development, testing, training, and mission operations. The NOS$^3$ environment contributed to the success of STF-1 mission in several ways, such as reducing the mission's reliance on hardware, increasing available test resources, and supporting training and risk reduction targeted testing of critical software behaviors on the simulated platform. The NOS$^3$ has been released as open-source and is available at http://www.nos3.org.

1. Introduction

   The NASA Independent Verification and Validation (IV&V) Program's mission is to provide assurance that safety- and mission-critical software will operate reliably and safely. NASA IV&V provides this service by employing a set of documented technical methods to the customers' system and software requirements, design, code, and tests. In 2009, the NASA IV&V Program established a simulation development team, the Independent Test Capa-bility (ITC). The ITC team is responsible for developing and maintaining test environments that are capable of exercising mission and safety critical software. IV&V teams are able to gain an increased understanding of the software execution and behaviors, exercise the system under adverse conditions, and inject faults into the system to gain insight into how the software will respond using ITC simulation environments. This capability thus enables the NASA IV&V Program to perform more thorough analyses of unit, build, and system level software tests and operational test procedures.

Since its inception, the ITC team has observed the benefits of software-only simulation environments to the IV&V Program and its customers but has also witnessed firsthand the benefits to software development organizations. ITC-developed software-only simulation environments have enabled risk reduction testing, provided earlier execution of operational tests, reduced the development organization's reliance on hardware, and increased available test resources on large spacecraft missions such as Global Precipitation Measurement (GPM) and James Webb Space Telescope (JWST). In addition to these large missions, the ITC team has applied its technologies to small satellites, which suffer from some of the same challenges such as long hardware lead times and software development/testing resources.

### 1.1 NASA CubeSat Launch Initiative

The NASA CubeSat Launch Initiative (CSLI) provides low-cost access to space for small satellites developed by NASA Centers and programs, educational institutions and non-profit organizations. NASA's investment in such technology is two-fold. First, the small satellite platform provides advanced educational opportunities for students, teachers, and faculty to help attract and retain students in Science, Technology, Engineering and Math (STEM) disciplines. Second, CSLI promotes partnerships between institutions to develop and mature low-cost technologies and pathfinders for the benefit of NASA programs and projects. Since its inception, the CSLI has selected 152 small satellite missions from 85 unique organizations. However, despite the increase of small satellite opportunities made available through CSLI, there remains a considerable amount of risk to these missions. Most of the standard risks involving cost and schedule apply, and are amplified, when dealing with the small-scale and fast-paced environment.

### 1.2 Simulation-to-Flight 1

As a result of the demonstrated successes of software-only simulation environments and the opportunity to launch a spacecraft to demonstrate technologies that benefit NASA programs through CSLI, the NASA IV&V Pro-gram and West Virginia University (WVU) collaborated to develop a 3U CubeSat mission, Simulation-to-Flight 1 (STF-1; Morris et al. 2016). The primary purpose of STF-1 was to determine and demonstrate the value of developing, utilizing, and maintaining a software-only simulation during the project lifecycle. However, a diverse set of science experiments, provided by WVU, allowed the project to expand the mission's overall objective. The instruments include a cluster of Micro Electro-Mechanical Systems (MEMS) Inertial Measurement Units (IMU) to produce attitude knowledge (Greenheck et al. 2014); a space-weather experiment including a Geiger counter and Langmuir probe (Vassiliadis et al. 2014); a III-V Nitride-based materials optoelectronics experiment (Pachol et al. 2016); and a Novatel OEM615 GPS coupled with advanced algorithms for precise orbit determination (Watson et al. 2016). The science experiments enhanced the mission capabilities as well as provided a diverse set of instruments to assess how the simulator would support science instrument development. Figure 1 provides an illustration of the various components and subsystems of the STF-1 CubeSat mission.

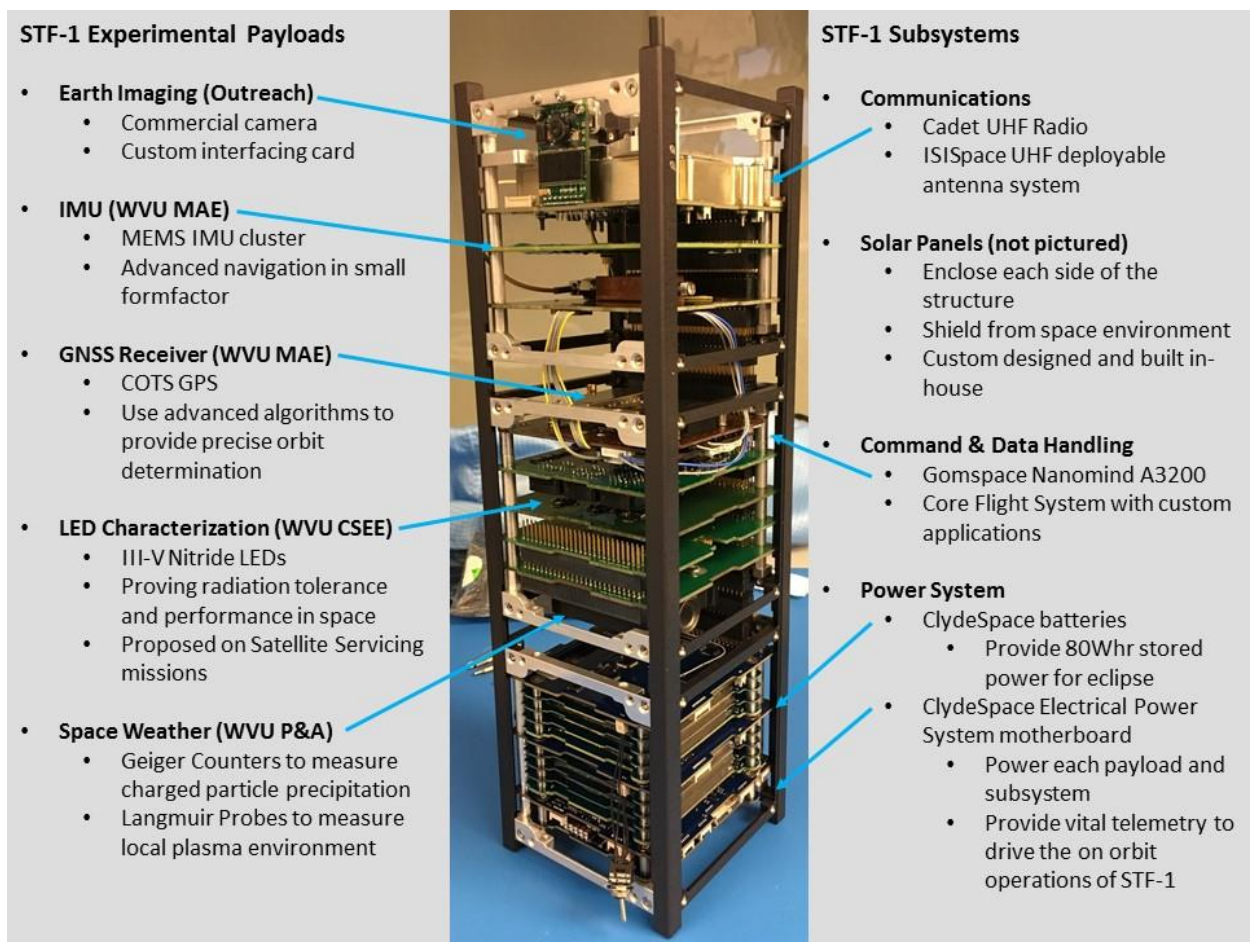

**Figure 1**. View of the components and subsystems of the STF-1 Cubesat, all of which have been simulated in NOS[3].

### 1.3 NASA Operational Simulator for Small Satellites

The STF-1 mission resulted in the development of a software simulation framework named the NASA Operational Simulator for Small Satellites (NOS[3]). The goal of NOS[3] is to enhance small satellite software development, testing, and training. With NOS[3], the flight software executes as if it were operating in space. NOS[3] provides the flight software with representative real-world simulated data inputs that it would expect during nominal on-orbit operations. Some of NOS[3] features include:

- enabling multiple developers to build and test flight software with simulated hardware models;
- serving as an interface simulator for science instrument / payload teams to communicate with prior to hardware integration;
- supporting software development activities;
- enabling hardware integration to parallel software development;
- providing automated testing framework;
- increasing available test resources and;
- enabling operation of the simulated spacecraft using the ground software command and telemetry databases.

### 2. NOS³ overview

An in-depth analysis of the NOS[3] and of its supporting products is presented in the following four subsections. Section 2.1 describes the high-level simulator architecture including software interfaces, simulated hardware models, and actual flight hardware. Section 2.2 describes the set of software tools that support the NOS[3] simulation architecture. These software tools consist of the NASA Operational Simulator (NOS) messaging middleware (NOS Engine), the open-source "42" general purpose multi-body, multi-spacecraft dynamic simulation (Stoneking 2008), the open-source COSMOS User Interface for Command and Control of Embedded Systems (Melton 2016), and the open-source core Flight System (cFS; Wilmot 2005), a platform- and project- independent, reusable software framework inclusive of a set of reusable software applications. Section 2.3 examines additional ad-hoc software that was developed to support and complement NOS³. Section 2.4 explores how NOS³ is deployed in a ready-to-run environment.

## 2.1    NOS³ Simulator Architecture

The flexible configuration of the NOS³ simulation architecture as compared to a typical flight system is illustrated in Figure 2. The "Flight Configuration" column provides a typical small satellite flight configuration for the flight software: i.e., flight applications, flight libraries, drivers, and flight hardware. The flight software may use flight libraries which provide common functionality. The flight software and libraries utilize hardware drivers, defined as software components communicating directly with the hardware. This is usually accomplished via reading and writing hardware registers and often using a bus protocol such as Universal Asynchronous Receive and Transmit (UART), Inter-Integrated Circuit (I2C), and Serial Peripheral Interface (SPI), or simply applying general purpose input/output (GPIO) signals.

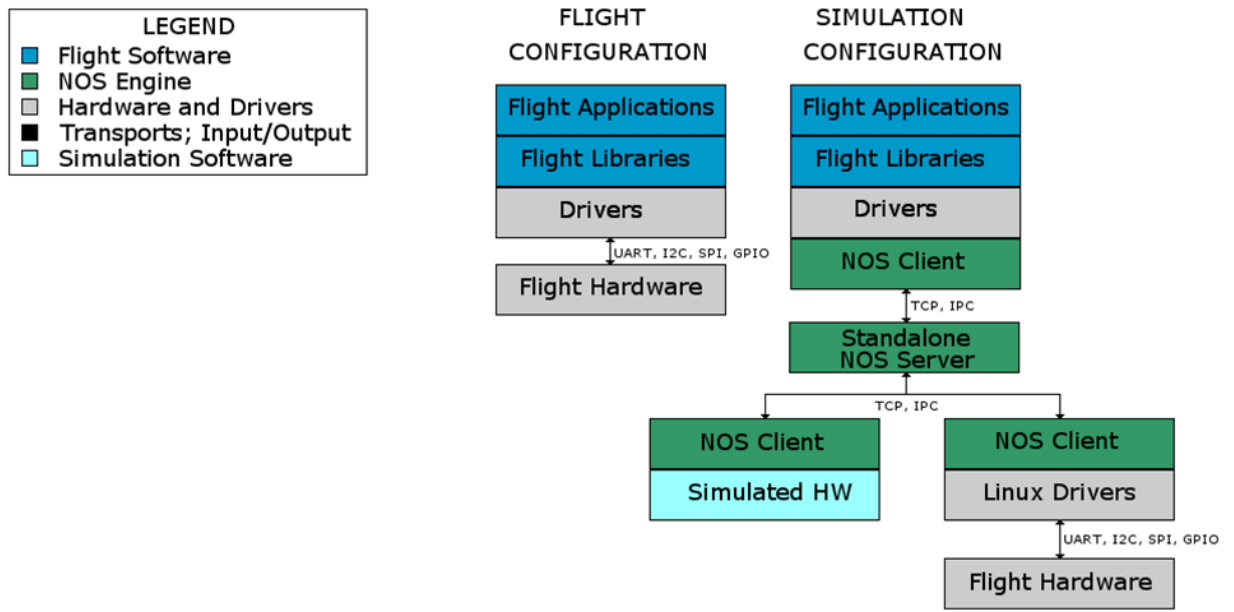

**Figure 2**. NOS³ Architecture. NOS³ architecture illustrating both its Flight and Simulation Configuration. Note that the Simulation Configuration is identical to the Flight Configuration with the exception of the interface to flight, or simulated Flight Hardware.

A typical satellite system has numerous hardware interfaces controlled through an on-board computer. These may include hardware interfaces with electrical power systems, radio frequency communication systems, science experiment payloads, orbit and attitude sensing and control systems, and others. The goal of NOS$^3$ is to substitute simulations in place of some or all of these hardware components.

The "Simulation Configuration" column demonstrates how NOS$^3$ can be utilized in place of the actual hardware. It should be noted that the NOS$^3$ architecture provides users with the flexibility to execute flight software with some or all of the hardware components replaced by a software simulation. This substitution occurs at the functional call interface. Performing the substitution is as simple as linking the flight software against a NOS$^3$ library to replace the hardware driver library. NOS$^3$ utilizes a client-server architecture and as such, a standalone NOS$^3$ server manages the communications between flight software and various hardware components. The standalone server maintains the components, referred to as nodes, that are attached to each hardware bus, the communications protocol used, etc. Additionally, NOS$^3$ includes a logging mechanism so that communications between simulation components can be monitored in real-time or in post-analysis to ensure that the data is passed correctly.

The hardware components that are being substituted with software simulations can be modeled at the fidelity required for the tests being performed. Some of the simulators written for STF-1 simply implemented pre-packaged data responses to commands from the on-board computer, while others required knowledge of the environment or other hardware components. For example, a GPS simulator will need to know the spacecraft position in orbit, therefore, this data must be generated dynamically. Simulators requiring this type of dynamic data utilize a connection to the "42" software (see Section 2.2.4) to collect the necessary data, and then proceed to package the response in the proper hardware format. In addition, the simulated components are able to be manipulated by the user, allowing fault testing that typically is not possible, or too dangerous to attempt in a hardware-only test.

**2.2 NOS$^3$ Software Components**

NOS$^3$ integrates a set of existing open-source software components as well as ITC developed software components to create a full spacecraft simulation platform. Figure 3 illustrates how these software components are interconnected within NOS$^3$. The following subsections examine each of these software components and their respective purpose: i.e., NOS Engine, COSMOS, cFS Flight Software, "42" Dynamics Simulator, Hardware Simulations.

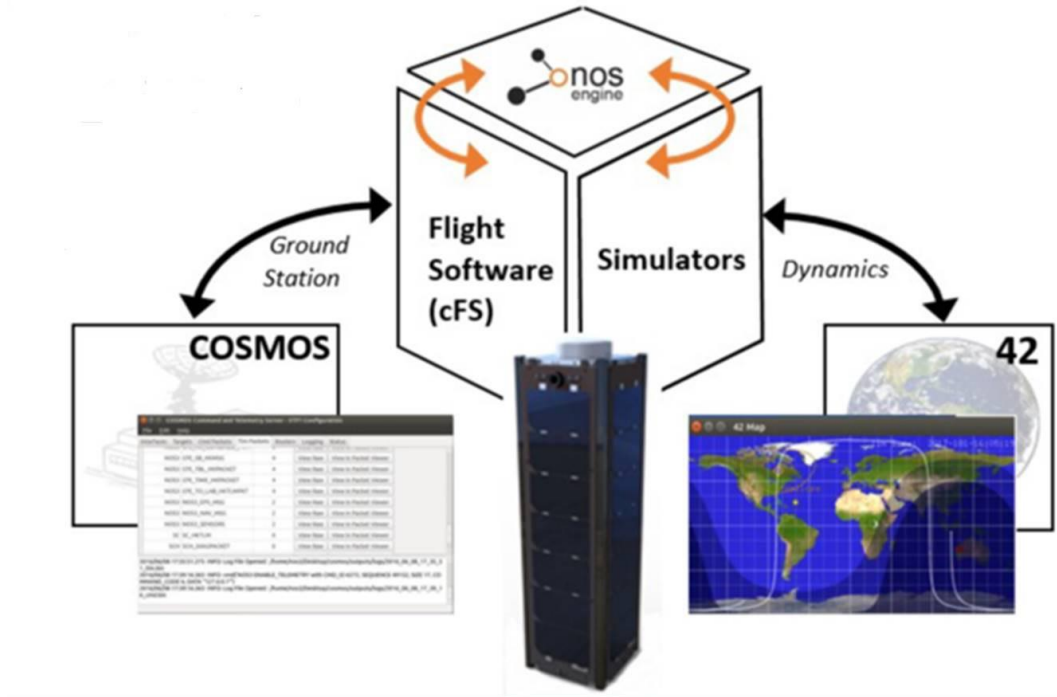

**Figure 3**. NOS³ components showing connections between ground station software (COSMOS), Flight Software (cFS), and Simulators/Dynamics (42).

### 2.2.1 Simulation Middleware: NASA Operational Simulator Engine

One of the primary software components of the NOS³ simulator is the NOS Engine simulation middleware that abstracts the hardware and connects the flight software with the simulated dynamics. NOS Engine is an in-house developed software suite that provides a library of functions to simulate the hardware communication protocols that are utilized by the flight software. As discussed in the previous sections, the hardware driver libraries are replaced with NOS Engine libraries utilizing the same function calls. NOS Engine also provides support for various underlying protocols such as TCP/IP, inter-process communication protocol (IPC), and shared memory to transport software bus messages that represent the actual hardware bus communication. This functionality provides a number of unique advantages: extremely fast communications; shared memory on a single computer running the flight software and the software simulators; and distributed processing such as TCP/IP on multiple computers, one running the flight software and others running the software simulators or interfacing with flight hardware, and other configurations based on development and testing requirements.

One of the challenges of simulated communications protocols (e.g., UART, I2C, SPI, etc.) is being able to represent their hardware time synchronization clocks within a software-only environment. Time synchronization clocks are used on small satellites for coordinating spacecraft time with ground time, coordinating time between various spacecraft components such as the on-board computer software and the radio frequency communication component, and providing timing signals for clocks that coordinate communication using protocols such as I2C and SPI. To overcome such a challenge the NOS Engine library contains methods to manipulate and distribute time between various components which are connected via software busses in place of what would normally be hardware busses. For example, within NOS$^3$, NOS Engine is utilized to control epochs and periodic clock signals.

### 2.2.2 Ground System Software: COSMOS

COSMOS (Melton 2016) is an open-source command and control software package and it was integrated into NOS$^3$ to allow end-to-end testing of STF-1 and to enable the "test as we fly, fly as we test" philosophy. COSMOS provides a sophisticated framework for command and control of satellites and other embedded systems. COSMOS was integrated into NOS$^3$ using a collection of text configuration files. A single text file provides the TCP/IP socket configuration information, while additional text files are auto-generated to define the byte patterns representing telemetry and command data sent from the spacecraft to the ground and vice versa.

NOS$^3$ includes several COSMOS enhancements to automatically generate and keep the data descriptions in the embedded code synchronized with the data descriptions in the COSMOS command and telemetry files. Data analysis mechanisms, in addition to what is provided with COSMOS, were required for the STF-1 mission and have been built as Ruby language extensions to COSMOS. These extensions are also available in NOS$^3$ and provide some of the post-processing data reduction for STF-1. It should be noted that despite COSMOS being already integrated into the NOS$^3$ framework, it is not architecturally required, and could be replaced by a similar command and control software that supports UDP connection.

### 2.2.3 Flight Software: core Flight System

The NASA-developed core Flight System (cFS; Wilmot 2005) is an open-source solution for spacecraft flight software, with flight heritage on numerous large and small NASA missions such as the Global Precipitation Measurement (GPM) and the Lunar Atmosphere Dust and Environment Explorer (LADEE). The cFS application layer includes a set of reusable software applications to support flight software development. The reusable applications are tailored to the mission requirements using tables, while new applications can also be developed for any mission specific requirement that is not directly provided by cFS. The software supports table driven applications, allowing applications to be tuned or changed during development and at runtime, by simply changing the tables' values without changing the code base. Another cFS component is a set of common services named the Core Flight Executive layer, that are typically needed by satellite systems such as time keeping and timers, executive services for applications, software bus messaging, and event reporting services. cFS is run on top of a lower level operating system framework called the Operating System Abstraction Layer (OSAL; Yanchik 2007), which isolates embedded software from the real-time operating system by providing users with an Application Program Interface (API). OSAL libraries are available for a range of operating systems including Linux, which allows NOS$^3$ libraries to be substituted at build time without any changes to the other cFS's layers. The Platform Support Package (PSP) is the cFS component that provides the interface to the hardware drivers for a specific on-board computer. NOS$^3$ is capable of substituting PSP libraries thus allowing the cFS to use standard function calls for various protocols (e.g., UART, I2C) to effectively communicate with the software simulations. The STF-1 mission and, therefore, NOS$^3$ made use of cFS not just for its flight heritage reliability, but also for this ability to substitute libraries that share a common API used by the flight software. It should be noted that it is architecturally possible to use NOS$^3$ without using cFS and OSAL. If cFS is not used an interface library would need to be written to utilize the NOS Engine API.

**2.2.4 Flight Dynamics: "42"**

A fundamental consideration in developing a small satellite simulator is how to provide realistic hardware signals reacting to the dynamically changing spacecraft environment. Specifically, as the spacecraft travels, variables such as its position, velocity, orientation, solar radiation direction and intensity, magnetic field direction and intensity change over time. While the actual hardware signals corresponding to dynamic inputs can be determined from hardware data sheets and user's manuals, the dynamic inputs must also be identified for a correct simulator development.

To provide a complete framework for spacecraft simulation, including the specific hardware simulations needed for the STF-1 project, we carried out a comprehensive analysis of different dynamic environmental data providers within NOS$^3$. After a thorough evaluation of numerous external solutions as well as the possibility of in-house development options, we chose the "42" software – a general purpose, multi-body, multi-spacecraft simulation – to provide dynamic environmental data (Stoneking 2008). "42" is an open-source software solution that provides the ability to propagate and predict the orbit and orientation of spacecraft, by computing the forces affecting these or-bital parameters, secondary gravitational effects, aerodynamic drag, solar radiation pressure, magnetic field interaction, and others.

### 2.2.5 Hardware Simulations

Several simulators have been developed for the hardware components utilized on STF-1, such as the GPS receiver, the antenna deployment system, and the electrical power system. While these simulators have features that are specific to the hardware components used on STF-1, they also present several elements useful to other satellite developers. For instance, they provide detailed, practical examples showing how simulators can be written for hardware components, how to use the NOS Engine communication busses, and how to receive dynamic data from "42". Furthermore, NOS$^3$ supplies a common simulation development framework for adding custom mission simulators; it includes functionalities for logging and text file configuration of simulators, it facilitates integrating custom mission capabilities and it assists with integrating environmental data providers such as "42". The framework also allows the user to create software simulators of a hardware component, early in the mission lifecycle to support flight software development and testing. These simulators can be written by referencing hardware interface control documents (ICDs) or data sheets, and further augmented with characteristic data from the hardware when available.

## 2.3 NOS$^3$ Supporting Software

In additional to the core NOS$^3$ simulation components, several other software components were developed and are included with NOS$^3$ to provide a more complete environment for operational use. Two of these components are described in the following subsections.

### 2.3.1 Mission Planning Software

An important part of satellite operations is mission planning. For satellite systems, it can include a multitude of tasks such as ground contact planning, power planning, planning when science data collection will take place, as well as data reduction once data is returned from the satellite. In the case of STF-1, ground contact planning was the primary procedure that needed to be formulated. In particular, when a ground contact takes place, the STF-1 operations team must be in communication with the radio antenna team at the antenna site to ensure that the proper data paths are configured and de-configured for the contact, and that the commanding and telemetry receipt planned for the contact time is planned and executed as quickly and efficiently as possible. Prior to the contact time, necessary personnel at the antenna and the STF-1 operations sites need to be adequately reserved to avoid scheduling conflicts. For STF-1 and other small satellites, ground contact occurs a few times per day with a typical duration of just a few minutes. The specific occurrence of these contacts can be accurately predicted using well-understood concepts of orbital mechanics coupled with a satellite's orbital elements, such as the two-line element (TLE) sets prepared by the United States Air Force and the North American Aerospace Defense Command.

NOS$^3$ provides a collection of Python utilities named Orbit, Inview, and Power Planning that can generate charts predicting accurate satellite visibility times from any location on Earth and when a satellite is in sunlight, Earth penumbra, and Earth umbra. The tools use TLE sets as their source of satellite orbital elements, and can generate tabular ephemeris data with rows indicating date and time, sub-satellite location on the earth, and satellite altitude. These data can be used for post-processing satellite data to correlate sensor observations with satellite position. For STF-1, science data such as radiation counts from the Geiger counter and plasma field data will be correlated with satellite position during post-collection data reduction activities.

### 2.3.2 NOS$^3$ Unit Test Framework

The benefits of unit and integration-level tests are well known, providing confidence that developed software operates as intended and future code changes do not cause unforeseen errors in other parts of the system (regression testing). The realization of the importance of testing early and often led us to include mature unit test frameworks for both the flight software and the simulators. We adopted the Google GTest framework for the NOS$^3$ simulators and the NASA UT-Assert library for the STF-1 flight software. The latter is the standard unit test framework for OSAL and core cFS applications.

Various additions and improvements were made to the UT-Assert library to simplify usage, such as integration into the build system and custom-built macros to simplify the process of creating unit tests. In addition, we created build targets for the GNU coverage testing tool and the Linux Test Program extension graphical front-end to allow the team to generate coverage reports and identify risk areas to improve testing. The separation of the hardware library in the STF-1 flight software architecture allowed separation of testing at both the application and hardware levels. For example, we tested applications with high-level inputs (commands, software bus messages, etc.), while a framework was created to stub hardware calls and allow the tester to provide appropriate low-level bus data for detailed hardware library testing.

Another important improvement to the unit test framework was the ability to run the unit tests on the STF-1 development and/or flight boards. Although testing in the NOS$^3$ simulation environment has proven to be beneficial, executing tests on the target architecture helped identify additional problems prior to hardware testing, further reducing integration times.

The easy-to-use unit test framework allowed developers to write tests in parallel to application development, uncovering issues early in the development cycle. This process saved integration time, in addition to code review time, since many bugs were already resolved by the developer prior to reaching those stages of the life cycle.

**2.4  Ready-to-Run Virtual Machines**

The NOS$^3$ collection of software components is conveniently packaged as a ready-to-run virtual machine, reducing the overhead associated with installing and configuring each software component. NOS$^3$ can be distributed as an Oracle VirtualBox virtual machine image or as a collection of command scripts that are used to recreate and modify a virtual machine image. This allows users to have a common development and testing environment, further reducing risk to the mission. The standard guest operating system utilized by NOS$^3$ is Ubuntu Linux but the virtual machine can run using Oracle VirtualBox on Windows, Mac, or Linux computers.

**3.  Results: the STF-1 CubeSat**

At the time of writing this paper, the STF-1 CubeSat mission software development and testing has completed and the spacecraft has been delivered for launch in 2018. Section 3.1 provides an overview of the software complexity of the STF-1 mission, while the remaining sections, Sections 3.2 to 3.4, highlight the three major benefits of using NOS$^3$ that were witnessed on the STF-1 mission.

**3.1 STF-1 Software Complexity Overview**

As a metric to assess the overall software complexity, the Source-Lines-of-Code (SLOC) utility (SLOCCount) was executed against the STF-1 flight software. This utility measures the size of a computer program by counting the number of lines in the program's source code. Additionally, the results of the SLOC utility were used as an indicator of software size for the Constructive Cost Model, a procedural cost estimation model. Table 1 lists the STF-1 SLOC count, with the RTOS and drivers not included because they were vendor provided. Of the 132,000 total SLOC, 25% of the software was newly developed for the STF-1 mission. Using the Constructive Cost Model, SLOCCount estimates that the STF-1 applications take 8.25 person-months for development, but this metric does not take into account unit testing, integration testing, and access to flight hardware for testing, which are typically the bottlenecks for small satellites and space missions.

Operationally, the STF-1 flight software is not trivial due to its semi-autonomous on-orbit functionalities that are needed to perform science experiments, record science data, and transmit the data to the ground station during downlink periods of just a few minutes long. The flight software must be able to simultaneously provide the following core functionalities: 1) operate without interaction/commanding from the ground station; 2) it must be aware of its power level status for executing time-lapse science experiments; 3) it must start, stop, and pause experiments; 4) it is responsible for communicating with various STF-1 hardware components such as sensors, radio, camera, and the deployable antenna. This flight software complexity results in increased mission-risk with respect to development and testing schedule. This type of embedded hardware testing is not possible without hardware-in-the-loop availability with the full ground-system software.

**Table 1**. STF-1 Flight Software SLOC Counts. STF-1 contained 34K SLOC (24%) of newly developed software.

| Software Component | Description | SLOC |
|---|---|---|
| Core Flight System (CFS) + Platform Support Package (PSP) | GSFC reusable flight software framework | 50 K + 7 K |
| Operating System Abstraction Layer (OSAL) | GSFC reusable operating system abstraction layer API | 41 K |
| STF-1 Mission Specific Applications | Newly developed flight software | 34 K |
| TOTAL | | 132 K |

**3.2. Reduced Hardware Reliance**

NOS[3] enabled multiple STF-1 developers to work in parallel without monopolizing either a single simulator, engineering test unit, or spacecraft flight computer, thus reducing the STF-1 mission's reliance on hardware. For example, while one engineer was developing the electrical power system software, another engineer was developing the communications software. Neither engineer needed to utilize the hardware for their development and initial testing.

NOS[3] was utilized extensively by the STF-1 software development team for all aspects of flight software development and testing. Over the course of three person-months in which most of STF-1 software development was accomplished, each team member maintained their own NOS$^3$ virtual environment. The virtual environment provided realistic inputs and feedback to the flight software while under development.

Additionally, NOS$^3$ provided a suitable test environment to support STF-1 flight software integration testing. Similar to many other small satellite missions, the STF-1 mission hardware was expensive, limited in supply with few spares, and it needed to be configured and setup quickly to support testing. NOS$^3$ provided the ability to develop and test most flight software functionality without requiring a hardware-in-the-loop test configuration. Hardware is still needed to test certain performance and timing requirements. Without NOS$^3$, STF-1 developers would not have been able to develop and test software applications in parallel to these activities. As a result, it would have been very difficult to maintain the flight software development and test schedule.

**3.3 Reduced Risk and Provided a Living Training Package**

The effortless deployment process of the NOS$^3$ software allowed us to setup and configure a large number of medium fidelity simulation environments to cross-train personnel and to support risk reduction testing during the STF-1 software development. For example, NOS$^3$ was provided to multiple interns during the summer months to support mission understanding, static analyses, and additional software testing of custom STF-1 software applications. The additional simulation resources allowed the team to test how the various STF-1 software applications would respond to adverse conditions thus ensuring STF-1 software robustness. One of the most critical STF-1 software applications, the manager application, which is responsible for semi-automating the spacecraft operations was exhaustively tested using NOS$^3$. NOS$^3$ also allows the tester to introduce fault conditions that are too dangerous or expensive to test using hardware, which further reduced mission risk and raised confidence in the flight software.

**3.4 Improved the Software Development Schedule**

NOS$^3$ was able to increase the STF-1 development team's control of the software development schedule and to demonstrate how future software development effort schedules can be shifted ahead of the receipt of hardware components. Table 2 reports the lead times associated with the major STF-1 flight components as compared with the associated development time for the NOS$^3$ hardware simulator. It is evident that the level of effort required to develop a hardware-equivalent simulator for the STF-1 mission with NOS$^3$ was rather minimal. Furthermore, a NOS$^3$ hardware simulator can be scoped, planned (effort, simulator fidelity, etc.), and efforted whereas hardware lead times from vendors change and slip regularly. NOS$^3$ allowed STF-1 software development to begin as scheduled versus when the hardware arrived.

**Table 2**. STF-1 component lead time compared to NOS3 software simulator development time. By reducing lead time, flight software development can start earlier in the mission

| Hardware Component | STF-1 Lead Time | NOS$^3$ Sim. Development Time |
|---|---|---|
| Antenna Deployment System | 6 months | 2 weeks |
| Electrical Power System | 10 months | 3 weeks |
| GPS Receiver | 2 weeks | 2 weeks |
| Magnetometer | 6 months | 1 week |
| UHF Radio | 7 months | 1 month |
| Experimental Payload | 12+ months | 1 week |

**4. Conclusions**

The primary purpose of the STF-1 CubeSat mission was to develop a software-only simulation framework and supporting tools that would support STF-1 as well as support future small satellite missions. The resulting byproduct of STF-1 is an open-source, software environment named NOS$^3$. The NOS$^3$ architecture was designed to be flexible and allow multiple configuration and deployment options. NOS$^3$ conveniently packages a set of open-source tools (cFS, COSMOS, and "42" dynamics simulator) and a set of STF-1 specific hardware simulators to provide a virtual spacecraft environment that is easy to configure and deploy to end-users. NOS$^3$ has demonstrated its extreme value to the STF-1 mission by reducing hardware reliance, increasing available test resources, serving as a training and risk reduction platform, enabling parallel software development activities that shorten cycles and reduce developer costs, and alleviating schedule pressures due to slips in hardware component deliveries. We plan to continue NOS$^3$ development in order to support future missions. As other teams adopt NOS$^3$ for their missions, additional hardware simulators can be added to the NOS$^3$ simulation library. NOS$^3$ can currently be found at www.nos3.org, and is available under the NASA Open Source License. To inquire about the NOS$^3$ software, e-mail the team at support@nos3.org providing a brief project introduction.

**ACKNOWLEDGMENTS**

The STF-1 development team would like to thank the NASA IV&V Program, West Virginia University science teams, TMC$^2$ Technologies, West Virginia Space Grant Consortium, and West Virginia High Technology Consortium. Without these contributors, our team would not have been building STF-1, "West Virginia's First Spacecraft".

Thank you as well to the Goddard Space Flight Center Dellingr, Ice Cube, and Ceres CubeSat missions for their collaboration and support.**REFERENCES**

Greenheck, D.R. et al. (2014): Design and Testing of a Low-Cost MEMS IMU Cluster for SmallSat Applications, presented at the 28th Annual AIAA/USU Conference on Small Satellites, Logan, UT, USA, August, 2014.

Melton, R. (2016): Ball Aerospace COSMOS Open Source Command and Control System, presented at the 30th Annual AIAA/USU Conference on Small Satellites, Logan, UT, USA, August, 2016.

Morris, J. et al. (2016): Simulation-to-Flight 1 (STF-1): A Mission to Enable CubeSat Software-based Verification and Validation, presented at the 54th AIAA Aerospace Sciences Meeting. San Diego, California, USA, 2016, Paper 6.2016-1464.

Pachol, M. et al. (2016): LOCC: Enabling the Characterization of On-Orbit, Minimally Shielded LEDs, presented at the 30th Annual AIAA/USU Conference on Small Satellites, Logan, UT, USA, August, 2016, Paper SSC16-P3-05.

SLOCCount: Available at: https://www.dwheeler.com/sloccount/ (accessed October 16, 2017).

Stoneking, Eric (2008): A computer program called "42" simulates the attitudes and trajectories of multiple spacecraft flying in formation anywhere in the Solar System, in NASA Tech Briefs, December 2008, Document ID 20080048040.

Vassiliadis, D. et al. (2014): Undergraduate Student-built Experiments in Sounding-Rocket and Balloon Campaign, presented at the AGU Fall Meeting, San Francisco, CA, USA, December 2014, ED21C-3456.

Watson, R., Sivaneri, V., Gross, J. (2016): Performance Evaluation of Tightly-Coupled GNSS Precise Point Positioning Inertial Navigation System Integration in a Simulation Environment, in *Proceedings of the AIAA Guidance Navigation and Control Conference (GNC)*, San Diego, CA, USA, January 2016.

Wilmot, J. (2005): A core flight software system, in *Proceedings of the 3rd IEEE/ACM/IFIP International Conference on Hardware/Software Codesign and System Synthesis*, 2005, pp. 13–14.

Yanchik, N. J. (2007): Operating System Abstraction Layer, presented at Flight Software Workshop, November 5-6, 2007, Laurel, MD, United States, Document ID 20080040870.
16